\documentclass[prd,aps,twocolumn,nofootinbib]{revtex4}

\usepackage[english]{babel}
\usepackage[letterpaper,top=2cm,bottom=2cm,left=3cm,right=3cm,marginparwidth=1.75cm]{geometry}

\usepackage{amsmath}
\usepackage{graphicx}
\usepackage[colorlinks=true, allcolors=blue]{hyperref}
\usepackage[utf8]{inputenc}
\usepackage{color}
\usepackage{xcolor}
\usepackage{multirow}
\usepackage{array}
\usepackage{subcaption}

\usepackage{verbatim}

\def\be{\begin{equation}}
\def\ee{\end{equation}}

\begin{document}

%TC:ignore
%\detailtexcount{main} % to get number of words
%TC:endignore

\title{Multimessenger probes of the galactic to extragalactic cosmic-ray transition}
\author{Clément Prévotat$^{1}$, Rafael \surname{Alves Batista}$^{1, 2, 3}$, Dmitri Semikoz$^{4}$, Sergey Koldobskiy$^{5}$}

\affiliation{$^1$Sorbonne Université, CNRS, Institut d’Astrophysique de Paris, 98 bis boulevard Arago, 75014 Paris, France}
\affiliation{$^{2}$Sorbonne Université, CNRS, Laboratoire de physique nucléaire et de hautes énergies, 75005 Paris, France}
\affiliation{$^{3}$Universidade de São Paulo, Institituto de Física de São Carlos, Av. Trabalhador São Carlense 400, 13566-590, São Carlos-SP, Brazil}
\affiliation{$^4$Université de Paris Cité, CNRS, Astroparticule et Cosmologie, F-75013 Paris, France}
\affiliation{$^{5}$Sodankyl\"a Geophysical Observatory and Space Physics and Astronomy Research Unit, University of Oulu, 90014 Oulu, Finland}

%TC:ignore

\begin{abstract}
The transition energy between Galactic and extragalactic cosmic rays remains an open question. Thanks to the new generation of  experiments, observational windows are opening in PeV energies in neutrinos (IceCube, GVD and KM3NeT) and gamma rays (HAWC, LHAASO), complementing CR measurements in this energy range. This expanded dataset provides a more comprehensive framework for investigating the origins of CRs.

This study seeks to estimate the contribution of extragalactic cosmic rays interacting in the Milky Way to the intensity of secondary particles such as gamma rays and neutrinos. Cosmic rays are injected from the border of the Galaxy, interacting with magnetic fields, photon fields and gas, producing secondaries during their propagation, which we use to reconstruct the intensity at Earth. 

We conclude that diffuse gamma-ray flux measured by LHAASO at PeV energies is dominated by Galactic sources, despite extragalactic cosmic rays can dominate the measured cosmic-ray proton flux above $10^{17} \; \text{eV}$. 
\end{abstract}
%TC:endignore

\maketitle

\section{Introduction} 
\label{Sec:Intro}

Explaining the features of the cosmic-ray (CR) spectrum remains one of the major challenges in astroparticle physics. Among the outstanding questions is the nature of the transition between the Galactic and extragalactic CR components, for which several scenarios have been proposed. The first of the two main hypotheses associates this transition with the second knee, located at approximately around $10^{17} \; \text{eV}$ \cite{Akeno_SecondKnee, Spectrum_KascadeGrande, Auger_SecondKnee}. In this picture, the second knee is interpreted as the counterpart of the proton knee for heavier Galactic nuclei. This scenario is supported by composition measurements indicating a relatively heavy CR composition around the second knee that gradually becomes lighter at higher energies, suggesting the emergence of a different, likely extragalactic, population of CRs~\cite{ReviewKachelriess, Rafa_CRReview_2024,Coleman_reviewUHECR_2023}. 
Additional support comes from the direction and low level of anisotropy measured in this energy range~\cite{Giacinti_CRAnisotropyTransition_2012}. However, this interpretation does not naturally account for the origin of the ankle.
The second paradigm places the Galactic-to-extragalactic transition at the ankle, around $5 \times 10^{18} \; \text{eV}$ \cite{Auger_CRSpectrum_2025, TA_CRSpectrum_2025}. In this case, the ankle is naturally interpreted as the crossover between the highest-energy heavy Galactic CR component and the emerging light extragalactic component \cite{ReviewKachelriess}. A key challenge for this scenario is the lack of firmly identified Galactic accelerators capable of reaching such energies.

Because charged CRs are deflected by magnetic fields during their propagation, inferring their spatial distribution is challenging: CR measurements need to be complemented by measurements of neutral messengers in order to build conclusive scenarios. Gamma rays beyond PeV are efficiently produced by CRs interacting with gas, mainly through the decay of produced neutral pions, and through interactions of CRs with photon fields permeating the Universe, which also produce neutral pions. Neutrinos are mainly produced through similar interactions, via the decay of charged pions. At these high energies, in an environment similar to the Milky Way (MW) (in terms of gas density and magnetic field), the contribution of electrons (or positrons) to the production of UHE gamma rays is subdominant as they suffer significant synchrotron losses~\cite{ReviewSigl}.

The detection of gamma-ray and neutrino fluxes at higher energies is becoming possible with new generation of experiments: LHAASO detects gamma rays with energies above PeV~\cite{LHAASO_PeVGalacticGamma_2021}, has already measured the diffuse gamma-ray flux up to $1 \; \text{PeV}$~\cite{LHAASO_DiffuseGammaKM2A_2023} and the Auger and Telescope Array experiments have set limits on the gamma-ray flux at the highest energies~\cite{Auger_LimitGammasHECO_2022, Auger_LimitGammaSD_2023, Auger_limitGammasLowE_2025, Auger_LimitGammasHybrid_2024, TA_LimitGamma_2021}, while IceCube  detected of neutrinos produced in the Galactic Plane~\cite{IceCube_NeutrinosGalacticPlane_2023,IceCube:2026plr} up to $0.1 \; \text{PeV}$, and KM3NeT recently announced the detection with very high probability of an ultra-high-energy ($\sim 220 \; \text{PeV}$) astrophysical neutrino~\cite{KM3Net_UHENeutrino_2025}. These results, together with future observations, could provide valuable insights into the CR populations responsible for the production of these secondaries. Source-associated secondary emission can probe particle acceleration mechanisms, while the diffuse flux traces the Galactic CR sea and its spatial distribution, which cannot be directly measured by local CR observatories. 

The objective of this paper is to estimate the diffuse fluxes of both gamma rays and neutrinos, produced by an extragalactic population of CRs propagating in the Milky Way. Assuming different extragalactic CR spectra, the variation of secondaries production due to a change of the assumed shape of the CR spectrum is explored, and we compare simulated fluxes with the latest measurements and constraints. In \ref{Sec:Simulations}, we describe the CRPropa simulation set-up, and subsequently, in~\ref{Sec: CR_spectrum}, we briefly introduce the CR spectra that we considered. Finally, results are presented in \ref{Sec:Results} and we conclude in \ref{Sec:Conclusion}.

\section{Simulations}
\label{Sec:Simulations}
We use CRPropa~\cite{CRPropa3_2016, CRPropa3.2_2022, CRpropa3.3_2025} to simulate the propagation of particles within the MW, using the Boris-Push algorithm to solve their equations of motion. CRs are injected at time $t_0$ from a sphere of radius 35~kpc surrounding the MW. Their injection locations are randomly distributed over the surface of this sphere, with momenta directed towards the interior of the sphere. Their directions are sampled according to a Lambertian distribution, in order to get an isotropic arrival distribution at any point in the MW. 
We simulated five different CR species: protons, helium, carbon, aluminium and iron nuclei. The energy spectrum of the injected particles in simulation is initially proportional to $E^{-1}$, with energies ranging from $10$~PeV to $100$~EeV. 
The injected particles are propagated through the MW until they escape, meaning until they get out of a sphere of 35~kpc centered on the Galactic Centre, or until their energy decreases below 10~PeV.

%The reason for that is that we are interested in high energy secondaries, that cannot be produced by CR of lower energies, and that propagating low energy particles is computationally expensive. textcolor{red}{THIS CAN NOT BE REASON. REASON IS GALACTIC WIND ... DS}. 

To speed up simulations, particles that travelled for more than 20~Mpc are also discarded. Only a few CRs close to 10 PeV energies that remain trapped in the Galactic magnetic field are discarded based on this criterion. The effect on the production of secondaries and on the CR densities is negligible. \\\\

\subsection{Galactic Magnetic Field}
Particles propagate in the Galactic magnetic field as described by the ``base'' model of the UF23 ensemble of models~\cite{UF23_regular_2023}. It describes the regular Galactic magnetic field using three components, a spiral disk field, a toroidal halo field, and an X-shaped field. We also accounted for the turbulent magnetic field, considering a modified JF12 model as prescribed in~\cite{JF12Planck_2016}, with a Kolmogorov spectrum and a coherence length of $60 \; \text{pc}$. The description of the turbulent magnetic field in this model stops at $r = 20 \; \text{kpc}$, setting the magnetic field to zero outside this region. By doing so, a discontinuity is created at the border of the MW, especially close to the Galactic plane where the turbulent field is stronger. While this feature has no real impact when propagating Galactic CRs in the MW or when simulating extragalactic the highest energy CRs, we observed in our simulations a ring-like structure in the x-y plane of the Galaxy along its outer border, feature that appeared progressively for the lowest energy particles ($\lesssim 10^{17} \; \text{eV}$) and decreased as the energy of the injected particles increased. To remove it, we extrapolated this magnetic field up to $r = 35 \; \text{kpc}$, assuming a similar evolution of the magnetic field strength with radial distance as was found for $r < 20 \; \text{kpc}$ in~\cite{JF12Planck_2016}, and inject CRs from this distance, in order to avoid injecting low rigidity particles in too strongly magnetized regions. 

\subsection{Gas distribution}
The main channel for gamma rays and neutrinos production is hadronic interactions, i.e., CRs interacting with the interstellar gas.
For atomic and molecular hydrogen, the gas distribution is taken from the recent model described in~\cite{Söding_GasDensity_2025}, in which three-dimensional Gaussian processes are used to model correlations in the interstellar medium, yielding spatially coherent 3D maps of the Galactic gas. For each studied species, 8 different models were produced. Using those 8 models, they in addition provide an averaged map as well as the associated standard deviation map, for each species. We use those averaged maps, over which some uncertainties estimations were computed, in order to estimate the impact of the gas model uncertainties on the production of secondaries. The ionized hydrogen (not modelled in~\cite{Söding_GasDensity_2025}), is subdominant in the disk compared to the other hydrogen populations, but can dominate the other species in the Galactic halo. It was modelled according to the models introduced in~\cite{InnerGasModel_Ferriere2007, OuterGasModel_Ferriere_1998}. Finally, the helium distribution follows the total hydrogen one, with a density reduced by a factor 10~\cite{Ferriere_ISMCompo_2001}.

\subsection{Radiation fields}
We considered interactions with different radiation fields: the Cosmic Microwave Background (CMB), the Extragalactic Background Light (EBL), the Cosmic Radio Background (CRB), and the Interstellar Radiation Field (ISRF). In our simulations, the EBL is described by the model presented in~\cite{EBL_Saldana2021}. The ISRF, which encompasses both the stellar light and the light reprocessed by dust within the MW, is described in~\cite{ISRF_Robitaille2012} with a spiral arms structure; interactions with the ISRF, were simulated using the module introduced in~\cite{ISRF_DiMarco2025}. Finally, the model we use for the CRB was introduced in~\cite{URB_Nitu2021}.

\subsection{Interactions}
While propagating, CRs can interact with gas or radiation fields.
Considered processes include inverse Compton scattering, synchrotron radiation, electromagnetic pair production, electromagnetic triplet pair production, photodisintegration, nuclear decays, electron pair production, photopion production and hadronic interactions. Depending on the propagated nuclei and on the target photon fields, some of the above interactions were neglected for some simulations, based on the cross-section tables provided in CRPropa~\cite{CRpropa3.3_2025}. The implementation of hadronic interactions is based on AAfrag~\cite{AAfrag_Kachelriess2019, AAfrag_Kachelriess2023}, more details can be found in supplemental materials \ref{Sec:HI}. The energy losses of  electrons are dominated by synchrotron radiation, while the production of gamma rays and neutrinos is dominated by hadronic interactions (with a contribution from photopion production at the highest energies), Gamma-rays can be absorded during their propagation ; it occurs mainly at energies below $10^{17} \; \text{eV}$, via electron pair production with the CMB, with a minor contribution from the ISRF at the lowest gamma-ray energies. 
Regeneration of the gamma-ray flux via pair production and inverse Compton scattering is found to be negligible compared to the first generation of gamma rays, and hence was then neglected to speed up simulations.

Because of the softness of the CR spectrum, the main contribution to the gamma-ray and neutrino intensities comes from light CRs, as the energy budget available for the secondaries scales as $\frac{E_{CR}}{A}$, with $A$ being the number of nucleons composing the parent CR. This was observed in our simulations, as a consequence we then focus only on proton and helium simulations.

There are strong evidences for the presence of a Galactic Wind at the Galaxy level \cite{Review_GWind_ruszkowski2023}. While it is often studied as a possible source for CR acceleration \cite{bustard_GWind_2017}, it can also affect the efficiency with which CRs penetrate within the MW, as studied in \cite{Mertens_2018_GWind}. The rigidity threshold above which CRs start becoming insensitive to this wind depends drastically on the parameters used to model it, and those parameters are not well constrained \cite{bustard_GWind_2017}. As a consequence, this threshold varies at least between $1~PV$ and $100~PV$ \cite{bustard_GWind_2017, Mertens_2018_GWind}. We decided to inject CRs with energies greater than $10~\text{PeV}$, and consider the lowest energy part of our results as an upper limit.

\section{CR spectrum}
\label{Sec: CR_spectrum}

\begin{figure}[htb]
    \centering
    \includegraphics[width=\linewidth]{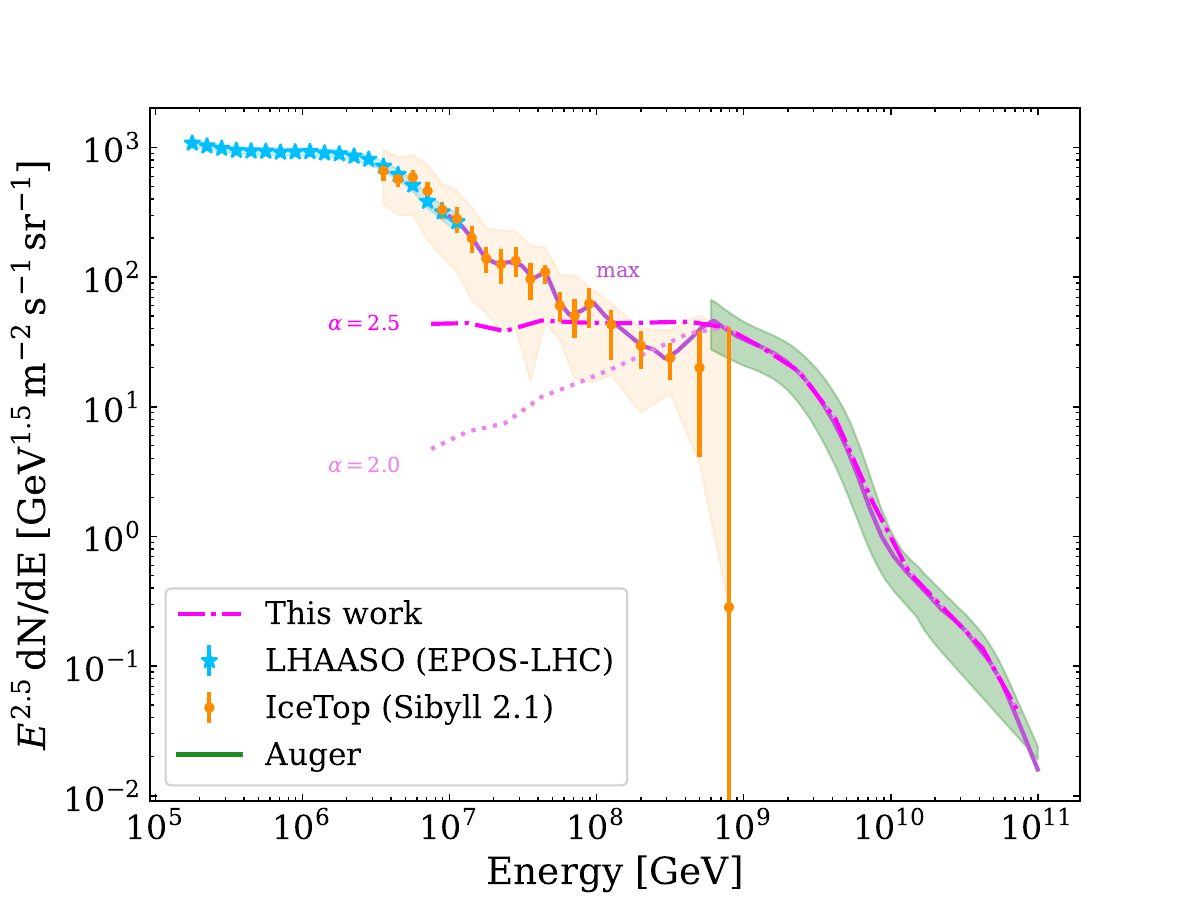}
    \caption{\label{fig:CR_spectrum} Extragalactic CR proton spectra are shown in different shades of purple, alongside with measurements by LHAASO~\cite{LHAASO_pHeSpectra_2025}, IceTop~\cite{Icetop_elements} and a model by Auger~\cite{Auger_BestFitSpectrum_2023}. }
\end{figure}

We inject approximately $3 \times 10^{9}$ protons and a similar number of He CRs in the MW.
Using the TimeEvolution module of CRPropa, each simulated particle is detected several times, each time after propagating for a given time period, corresponding to a travelled distance of $ \sim 0.67 \; \text{kpc}$ (for particles moving at the speed of light). 
The steady state distribution can be reconstructed from the distribution of particles in all the snapshots by considering the steady state to be the sum of the distribution at each snapshot. The distribution of proton (He) CRs accounts exclusively for protons (He) produced in simulations of protons (He), as those produced in simulations of heavier nuclei propagating within the MW give a negligible contribution.

The propagated cosmic-rays are reweighted according to their injection energy (or in case of secondary CRs, according to the injection energy of their parent CR) so that the intensity in the bin where the Earth is located matches the local intensity at high energies proposed as extragalactic contributions in~\cite{Auger_BestFitSpectrum_2023}. In this article (\cite{Auger_BestFitSpectrum_2023}), this model is proposed to explain the ankle as the superposition of two extragalactic components, one dominating at low energies and the other at high energies. As it describes only the high energy part of our simulated CRs (energies above 0.6~EeV), we reweight the lowest energy part of the CR spectrum injected at the border of the MW by a simple power-law, whose index is varied between $2.0$ and $2.5$. We also consider a maximal case, where the CR intensities at Earth at energies smaller than 0.6~EeV match the spectra reported by LHAASO, IceTop and Auger. The different CR spectra we considered are plotted in Figs. \ref{fig:CR_spectrum} and \ref{fig:CR_He_spectrum}, along with some experimental measurements.

\section{Results}
\label{Sec:Results}

Neutrinos and gamma rays are detected by a spherical observer of the size of the Earth and located at Earth. As expected, we find that the main channels at play for the production of gamma rays and neutrinos here are photodisintegration by CMB and EBL photons, and hadronic interactions with Galactic gas. Hadronic interactions dominate at most energies, but there is a non-negligible contribution from photopion production at the highest energies (above $\sim 3$~EeV). Moreover, the contribution of secondary protons is negligible because their parent CR have higher energy, so their flux is strongly reduced when reweighting particles. 

\begin{figure}[htb]
    \centering
    \includegraphics[width=0.95\linewidth]{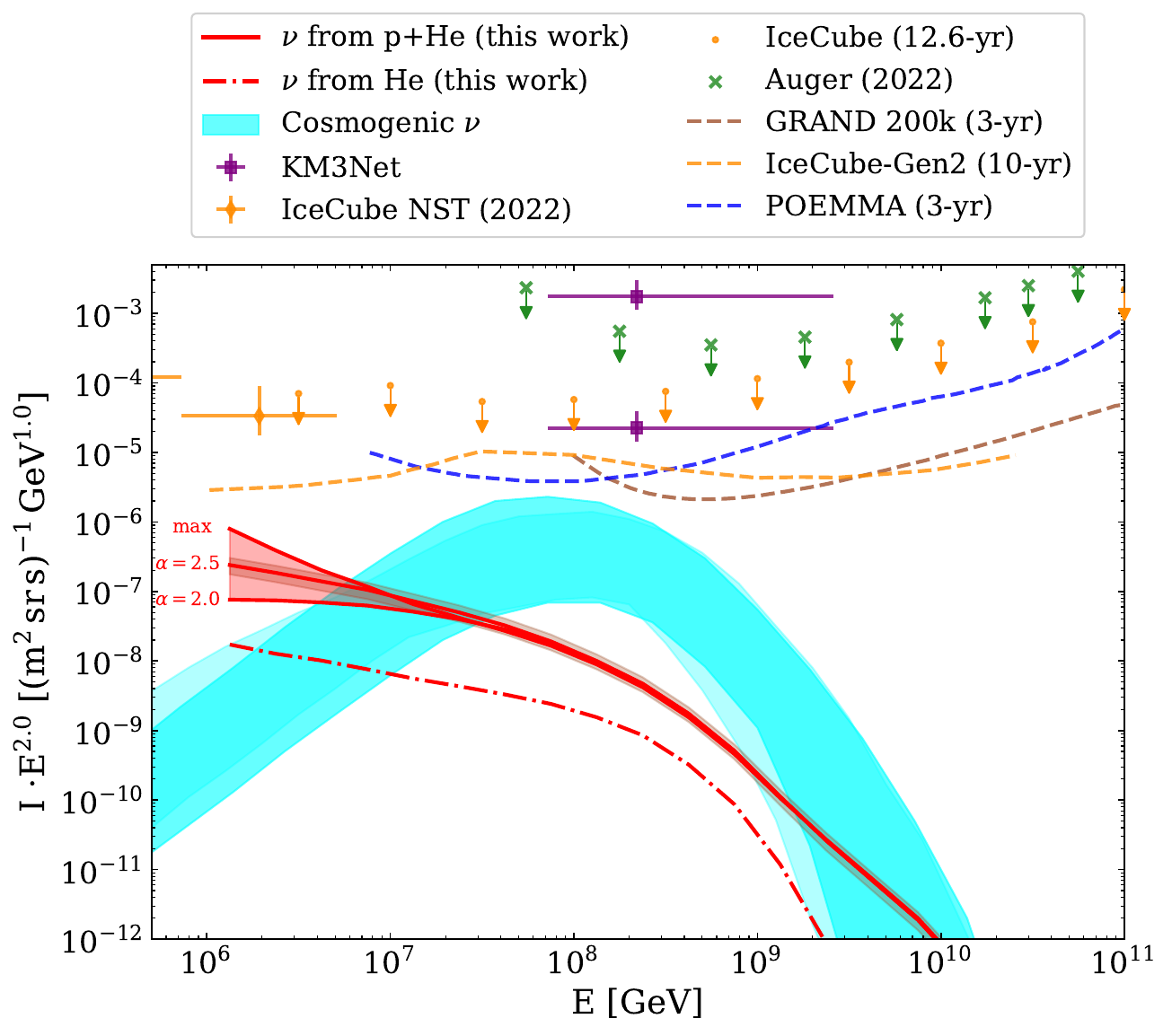}
    \caption{\label{fig:nu_limits} All-flavour neutrino intensity obtained in this work (solid red lines), compared with  cosmogenic neutrino model (cian shaded region)  \cite{Cosmogenic_GamNu_2019_Rafa},  the KM3NeT high-energy event \cite{KM3Net_UHENeutrino_2025, KM3NetEvent_in_GlobalNeutrinoLandscape}, IceCube data \cite{IceCube_neutrinos_points_2022}, limits obtained by IceCube~\cite{IceCube_NeutrinoLimits_2025} and Auger~\cite{Auger_DiffuseNeutrinosFlux_2022}, along with the sensitivity of some future experiments: GRAND \cite{GRAND_WhitePaper}, POEMMA \cite{POEMMA_2019} and IceCube-Gen2 \cite{Review_Ackermann_2022}.}
\end{figure}

 Nonetheless, we observe that depending on the composition assumed for the CR spectrum, a non negligible contribution from heavier CRs can arise : in our case, in figure \ref{fig:nu_limits}, the helium contribution to the neutrino and gamma-ray intensity becomes comparable to the proton contribution at $E \sim 1$~EeV, which is directly related to the helium bump we see in the CR composition model we use.

The mean intensity of cosmic rays and secondaries are reconstructed integrated over the entire sky. We also compute them over two regions of interests in the Galactic plane, the ones used in LHAASO measurements of the diffuse gamma-rays flux~\cite{LHAASO_DiffuseGamma_2025}: the so-called ``inner'' part of the Milky-Way, corresponding to $|b| < 5^\circ$ and $15^\circ < l < 125^\circ$, and the ``outer'' part, defined by $|b| < 5^\circ$ and $125^\circ < l < 235^\circ$. In this section, we compare the intensities we obtain for gamma rays and neutrinos with present measurements and limits. Our results can be compared with those of the study presented in \cite{joshi_2014_GCneutrinos}, in which simple geometrical models were used for the gas distribution and the CR density was assumed to be constant throughout the MW.

In Fig.~\ref{fig:nu_limits}, the obtained all-sky diffuse neutrino spectrum is compared with limits computed by the IceCube~\cite{IceCube_NeutrinoLimits_2025}  and Auger~\cite{Auger_DiffuseNeutrinosFlux_2022} collaborations. We also plotted the sensitivity of several future neutrino observatories, such as GRAND \cite{GRAND_WhitePaper}, IceCube-Gen2 \cite{Review_Ackermann_2022} and POEMMA \cite{POEMMA_2019}. The two purple points describe the high energy neutrino event detected by KM3Net, the upper one when considering only the KM3Net observatory \cite{KM3Net_UHENeutrino_2025}, the lower one when including in the fit of the spectrum the non-detection of similar neutrino events by IceCube and Auger \cite{KM3NetEvent_in_GlobalNeutrinoLandscape}. Finally, the gray band represents the $1\sigma$ uncertainty due to the gas models. Between 1~and 10~PeV, our flux lies around 2~orders of magnitude below the IceCube’s measurements, but only one order of magnitude below the sensitivity of IceCube-Gen2. When considering the maximal case, where the local CR spectrum follows local measurements, the all-sky neutrino intensity is only three times lower IceCube-Gen2’s sensitivity (for 10 years). At 10~PeV, this intensity is 2~orders of magnitude below POEMMA's sensitivity over 3~years, and constraints get looser as the energy increases. Those results are compatible with what was found in \cite{beratFloorCosmogenicNeutrino2024}, where only the highest energies were considered.

In order to compare our result with another astrophysical flux, we plotted the cosmogenic neutrino intensity obtained in~\cite{Cosmogenic_GamNu_2019_Rafa} considering a source evolution scaling in $(1 + z)^m$, where m is a free parameter, and $z_{max} = 1 (4)$ for the bright (faint) band. Interestingly, our results show that the two intensities are comparable in some energy ranges, though with different shapes: as our result is dominated by hadronic interactions, the shape of the resulting spectrum follows closely that of the parent CR spectrum, while the cosmogenic neutrinos come from photopion production with the CMB and the EBL, as well as photodisintegration, which yield a more bumpy spectrum. Those intensities could be distinguished by looking at the arrival direction of detected neutrinos: neutrinos produced in the MW originate mostly from within the Galactic plane, where most of the gas is concentrated, while cosmogenic neutrinos are expected to have a more isotropic distribution. 

\begin{figure}[htb]
    \centering
    \includegraphics[width=0.95\linewidth]{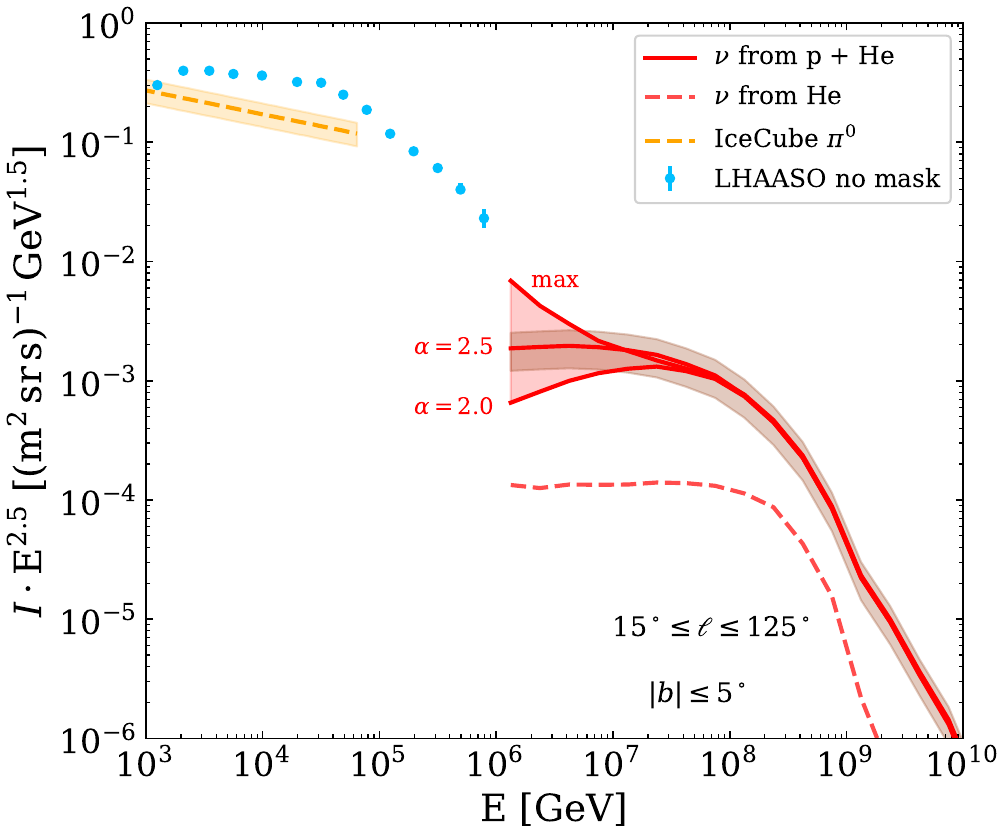}
    \caption{\label{fig:neutrinos_inner} Comparison of the diffuse all-flavour neutrino intensity in the inner region of the MW with IceCube’s measurements (rescaled to this sky region) \cite{IceCube_NeutrinosGalacticPlane_2023, Prévotat_LocalKnee_2025}.}
\end{figure}

We plot this neutrino intensity in the Galactic plane, both in the inner and outer regions in Fig. \ref{fig:neutrinos_inner} and \ref{fig:neutrinos_outer}, along with the $\pi^0 ~\text{Fermi}$ estimate produced by IceCube \cite{IceCube_NeutrinosGalacticPlane_2023} and rescaled to those sky regions \cite{Prévotat_LocalKnee_2025}. We also show LHAASO’s diffuse gamma-ray measurement as the production of neutrinos and gamma rays in hadronic interactions are tightly related \cite{muraseTestingHadronuclearOrigin2013}.

\begin{figure}[htb]
    \centering
    \includegraphics[width=0.95\linewidth]{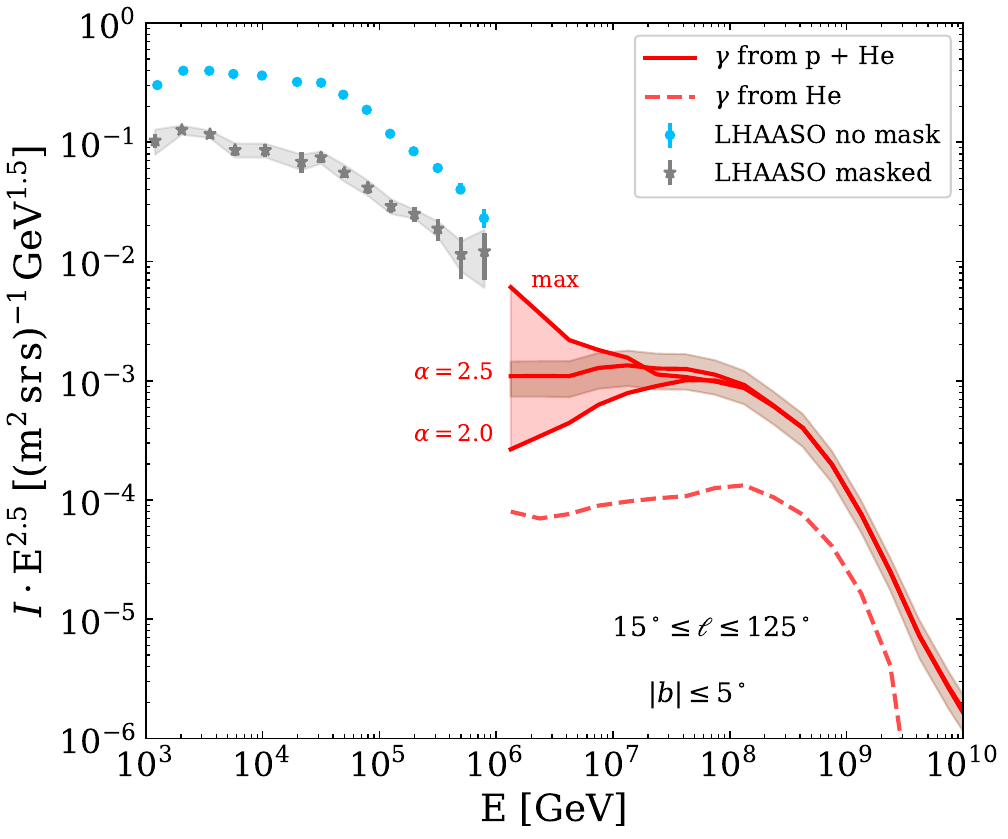}
    \caption{\label{fig:gammas_inner} Comparison of the diffuse gamma-ray intensity in the inner part of the MW with LHAASO’s measurements \cite{LHAASO_DiffuseGamma_2025}.}
\end{figure}

For gamma rays, we considered the recent measurements of the diffuse galactic gamma-ray flux reported by LHAASO~\cite{LHAASO_DiffuseGammaKM2A_2023}, both in the inner and outer regions, and upper limits set by the Auger~\cite{Auger_LimitGammaSD_2023, Auger_LimitGammasHECO_2022, Auger_LimitGammasHybrid_2024, Auger_limitGammasLowE_2025} and Telescope Array~\cite{TA_LimitGamma_2021}. 
The situation is slightly different: as shown in Fig.~\ref{fig:int_gammas_all_sky}, as in the case of neutrinos, the all-sky integrated gamma-ray intensity is much lower than present limits between 1~and 20~PeV, around 2~orders of magnitude lower, and is even lower at higher energies. However, when comparing our results in the Galactic plane (Fig.~\ref{fig:gammas_inner}), depending on the injected CR spectrum, the diffuse intensity as measured by LHAASO can be comparable to our results. While at 1~PeV, with an injected CR spectrum proportional to $E^{-2.5}$ at the lowest energies, our results are around 20~times smaller than LHAASO's measurements (without any mask), when injecting a CR spectrum from 1~PeV to 100~EeV which follows the one we measure at Earth, our result could be comparable with LHAASO future measurements around 2~PeV. This is especially the case when looking at the outer region (Fig \ref{fig:gammas_outer}), where we expect to have less high energy CR sources than in the inner region. This could point towards an extragalactic contribution to the diffuse gamma-ray flux, though this case is a stringent upper limit as it assumes a negligible contribution from Galactic CRs. %Also, one should keep in mind that the typical energy of CRs producing gamma rays through hadronic interactions is around 10~times higher than that of the produced photons: those CRs are protons beyond the knee, where Galactic production is thought to become sparser.of order

%This last point could also open some space for future works on constraining the extragalactic magnetic field (EGMF): depending on the intensity and coherence length of the EGMF, of the source distribution, and of the composition at the sources, the efficiency of the magnetic horizon effect~\cite{Deligny_magnetic_horizon_2004, aloisioDiffusivePropagationUltra2004, alvesbatista2014b} in depleting the proton (or He) flux in the MW at energies $\sim 10^{17} \; \text{eV}$ (typical energy range to produce gamma rays and neutrinos between 1~and 10~PeV) varies. In turn, those parameters directly affect the flux of secondaries produced by EGCRs propagating in the MW, which is what we computed in this paper, assuming a given EGCRs intensity at Earth. On the other hand, if the CRs that produced those secondaries originate from within our Galaxy, the magnetic horizon effect would not be relevant. In this case, we could however expect strong inhomogeneities in the spatial distribution of observed gamma-rays, depending on the sparsity of such powerful accelerators within the MW. 

\section{Conclusion}
\label{Sec:Conclusion}

 We computed the gamma-ray and neutrino diffuse intensity produced by different models of extragalactic CR spectra. While the neutrino flux is found to be out of reach of both neutrino experiments operating nowadays and the next generation of neutrino observatories, it remains comparable to the cosmogenic flux as computed in~\cite{Cosmogenic_GamNu_2019_Rafa} around 10~PeV and above~EeV energies. As the neutrino intensity we computed is dominated by the emission along the Galactic plane, this intensity would be scaled up when looking only at the Galactic plane (by a factor $\sim 10$), contrary to the cosmogenic neutrino intensity, for which there is no preferred direction of observation.

The situation is found to be somewhat similar for the all-sky gamma-ray flux, with a flux $\sim 2$~orders of magnitude below the most stringent limits in the best case. However, when looking specifically at the Galactic plane, the diffuse gamma-ray (and neutrino) flux is found to be much closer to LHAASO measurements, and even comparable, depending on the CR spectrum we assume for the extragalactic component. However, as at least 43 Galactic PeVatrons are already seen \cite{LHAASO_Catalogue2023}, the "maximal" case is a stringent upper limit over the extragalactic CR flux in the MW, therefore supporting that the diffuse gamma-ray emission seen by LHAASO is dominated by Galactic CRs. 

%The spatial distribution of the diffuse gamma-ray intensity above $1 \; \text{PeV}$ could then be used to disentangle and constrain the Galactic and extragalactic CRs population.

\vspace{-0.25cm}
%TC:ignore
\section{Acknowledgements}
RAB and CP acknowledge the support of the Agence Nationale de la Recherche (ANR), project ANR-23-CPJ1-0103-01. Work of DS was  supported in part under the program ``Investissement d'Avenir'' launched by the French Government and implemented by ANR, with the reference «ANR‐18‐IdEx‐0001» as part of its program «Emergence».
%TC:endignore
\vspace{-0.25cm}

\clearpage
%TC:ignore
\bibliography{refs}
%TC:endignore

\clearpage

%TC:ignore

\appendix
\section{He CR spectrum}
\label{Sec:He_CR_spectrum}

\begin{figure}[htb]
    \centering
    \includegraphics[width=\linewidth]{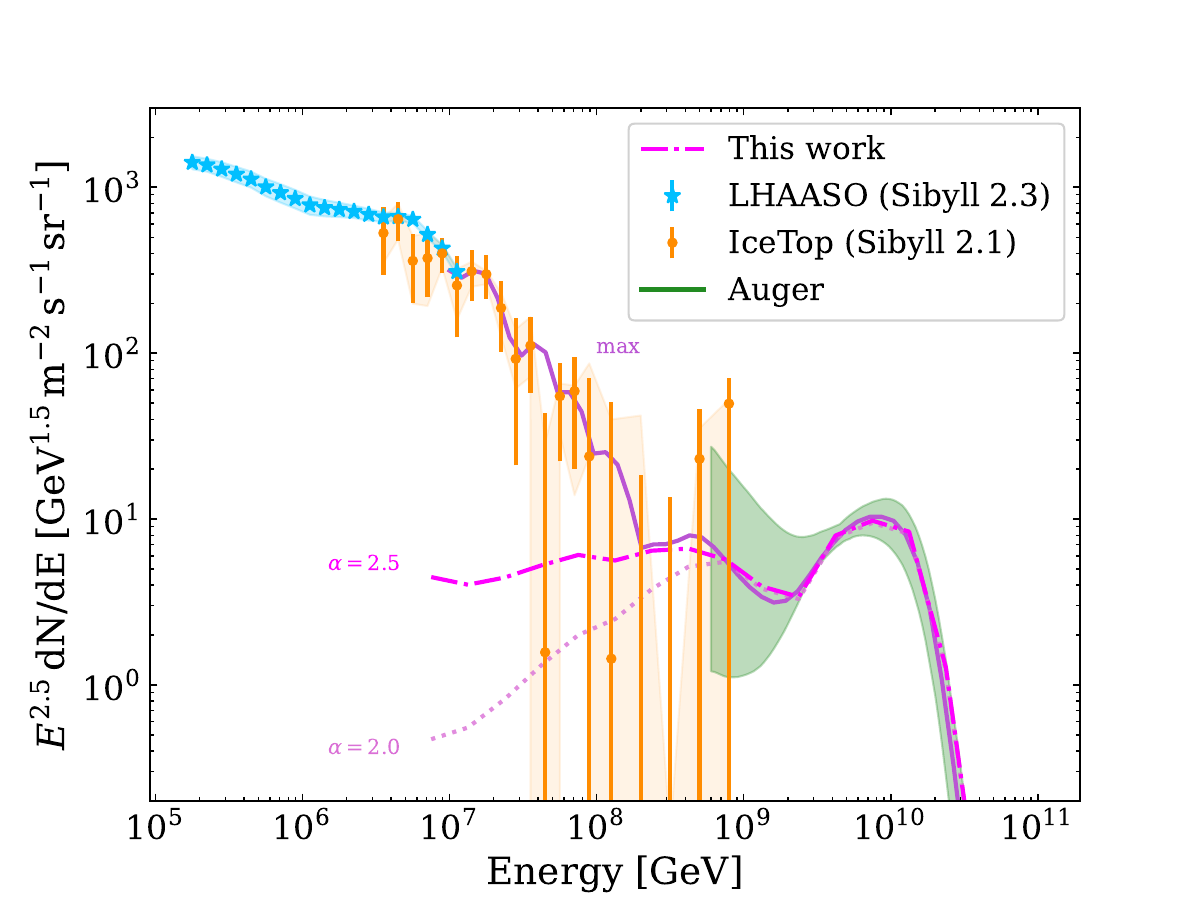}
    \caption{\label{fig:CR_He_spectrum} Assumed extragalactic CR He spectra are shown in different shades of purple, alongside with measurements by LHAASO~\cite{LHAASO_pHeSpectra_2025}, IceTop~\cite{Icetop_elements} and a model by Auger~\cite{Auger_BestFitSpectrum_2023}.}
\end{figure}

We plot in Fig.~\ref{fig:CR_He_spectrum} the different intensities we assumed for the extragalactic He component : Auger’s fit \cite{Auger_BestFitSpectrum_2023} above 0.6~EeV, and 3 different power-law like intensities at lower energies, $\alpha = 2.0$, $\alpha = 2.5$ and a maximal case that "follows" IceTop’s measurements \cite{Icetop_elements}. LHAASO’s He spectrum \cite{LHAASO_pHeSpectra_2025} is plotted for comparison. 

\section{Results in the "outer" region}

\begin{figure}[htb]
    \centering
    \includegraphics[width=0.95\linewidth]{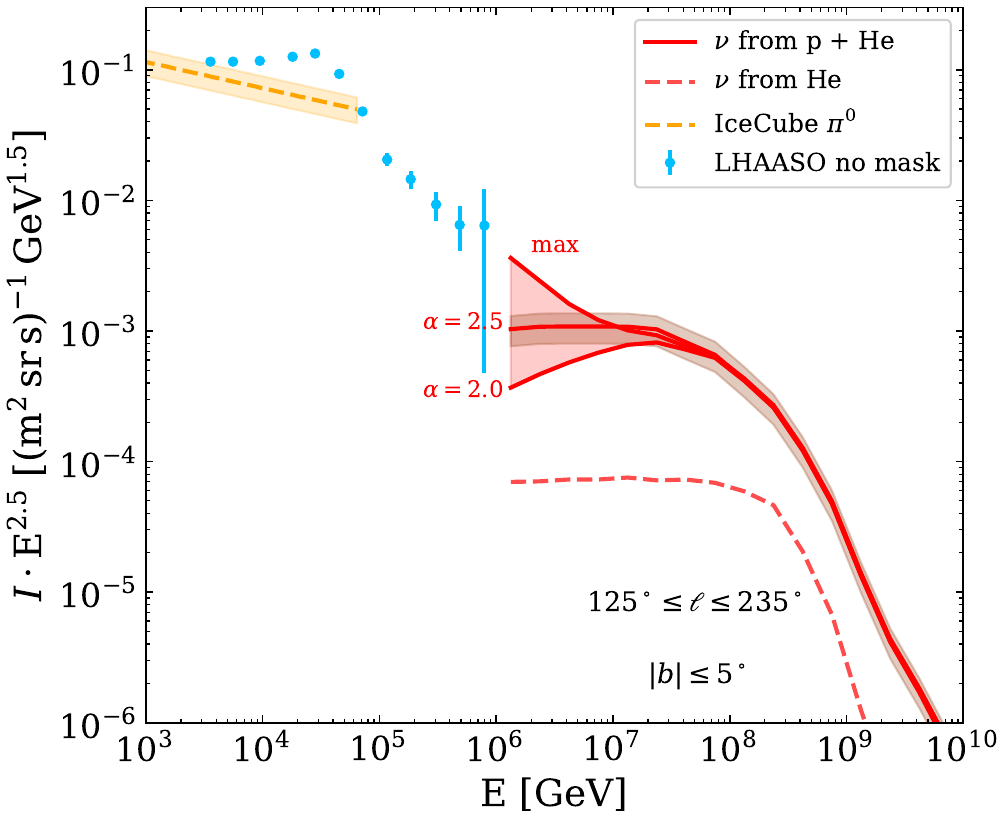}
    \caption{\label{fig:neutrinos_outer} Comparison of the diffuse all-flavour neutrino intensity in the outer region of the MW with IceCube’s measurements (rescaled to this sky region) \cite{IceCube_NeutrinosGalacticPlane_2023, Prévotat_LocalKnee_2025}.}
\end{figure}

\begin{figure}[htb]
    \centering
    \includegraphics[width=0.95\linewidth]{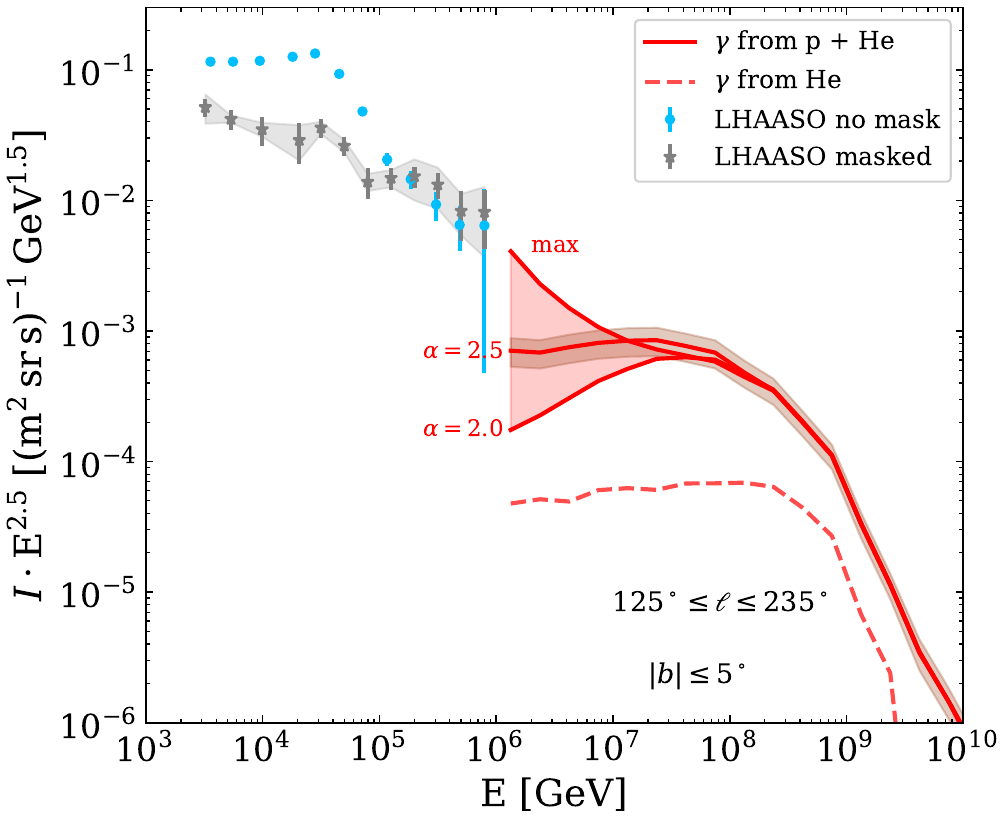}
    \caption{\label{fig:gammas_outer} Comparison of the diffuse gamma-ray intensity in the outer region of the MW with LHAASO’s measurements ~\cite{LHAASO_DiffuseGamma_2025}.}
\end{figure}

In Fig.~\ref{fig:neutrinos_outer} and \ref{fig:gammas_outer}, we show results of similar computations as in the main text, but for the outer part of the galactic plane. 

\section{Integrated gamma-ray intensity}

\vspace{-0.65cm}

\begin{figure}[htb]
    \centering
    \includegraphics[width=0.95\linewidth]{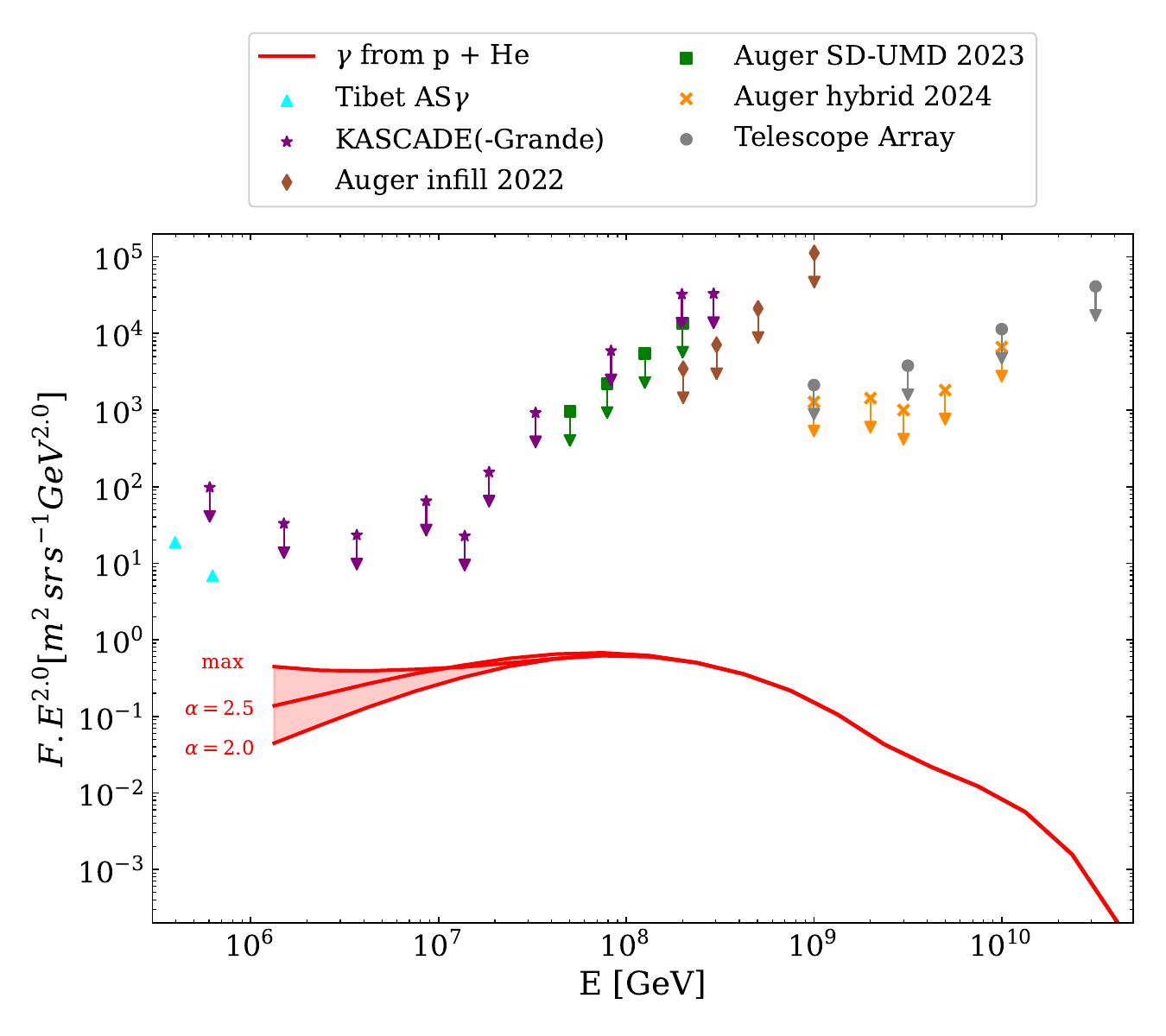}
    \caption{\label{fig:int_gammas_all_sky} Integrated all-sky gamma-ray intensity, and upper limits measured by Tibet-AS${\gamma}$ \cite{Neronov_TibetDiffuseLimit_2021},  KASCADE(-Grande)~\cite{KASCADE_GammasIntegrated_2017}, Auger~\cite{Auger_LimitGammasHybrid_2024, Auger_GammasIntegratedSDUMD_2023, Auger_LimitGammasHECO_2022} and Telescope Array~\cite{TA_LimitGamma_2021}.}
\end{figure}

In Fig. \ref{fig:int_gammas_all_sky}, we plot the integrated all-sky gamma-ray intensity, and compare it to the latest available limits, computed by Tibet-AS${\gamma}$ \cite{Neronov_TibetDiffuseLimit_2021}, KASCADE(-Grande)~\cite{KASCADE_GammasIntegrated_2017}, Auger (infill, surface detector and underground muon detector (SD-UMD) and hybrid (surface detector + fluorescence detector)) ~\cite{Auger_LimitGammasHybrid_2024, Auger_GammasIntegratedSDUMD_2023, Auger_LimitGammasHECO_2022} and Telescope Array~\cite{TA_LimitGamma_2021}. The diffuse flux we compute remains below those limits, for all the different CR flux that we assumed.

\section{Hadronic interactions}
\label{Sec:HI}

\begin{figure}[htb]
    \centering
    \includegraphics[width=1\linewidth]{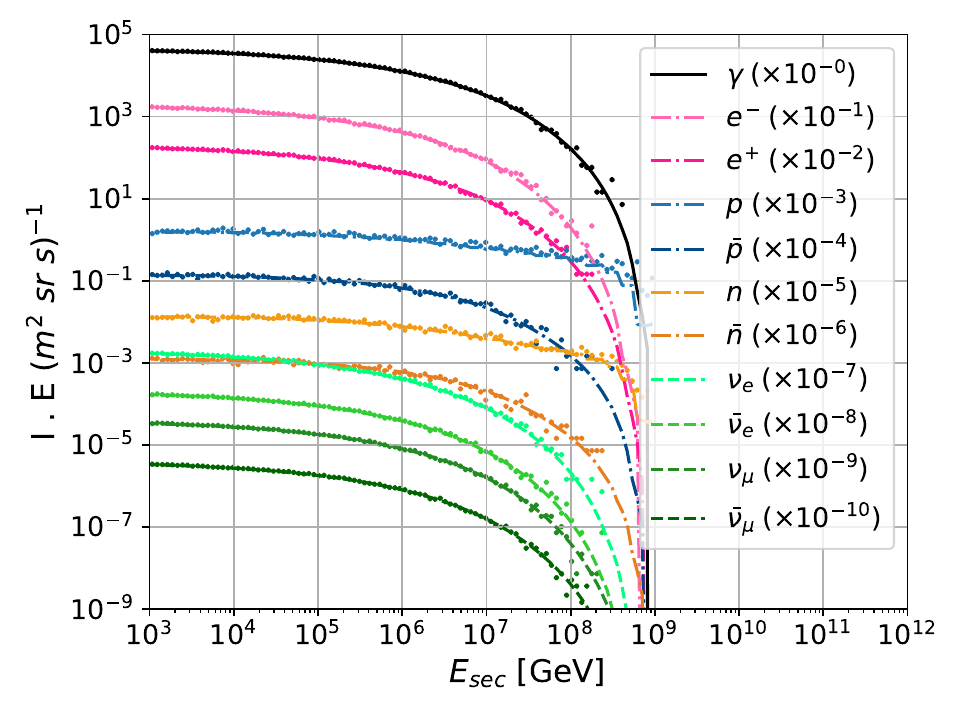}
    \caption{Spectra of secondaries, produced by protons, computed using AAfrag (lines) and from CRPropa simulations (dots).}
\label{fig:HI_comp_1_1}
\end{figure}

In our simulations, we included hadronic interactions between CR and Galactic gas, which we considered to be composed of hydrogen and helium as described in section \ref{Sec:Simulations}. Starting from the CRPropa plug-in introduced in~\cite{HI_Dorner2025}, we added interactions between $\text{He}$, $\text{C}$, $\text{Al}$ and $\text{Fe}$ cosmic rays and hydrogen gas, and interactions between protons or helium cosmic rays with helium gas. This was achieved using AAfrag202~\cite{AAfrag_Kachelriess2019, AAfrag_Kachelriess2023}, which tabulates the production cross-sections of gamma rays, electrons, neutrons, protons, electron and muon neutrinos, and all the corresponding anti-particles, in hadronic interactions, based on simulations ran with QGSJET-II-04m. Moreover, since the highest tabulated energies of the primary particles is $\sim 3 \; \text{EeV}$ (except for $p-p$ or $C-p$ interactions) we extrapolated to higher energies most cross-sections tables. This approach doesn’t pretend to model those ultra-high-energy interactions very precisely, but is rather seen as a way to coherently take into account such interactions. 

The total inelastic cross-sections are computed following the description in~\cite{Kafexhiu_CrossSections_HI_2014}. A comparison of simulations using our hadronic interaction module with the results given analytically by AAfrag is shown in Fig.~\ref{fig:HI_comp_1_1}. This comparison was made considering a spatially constant gas distribution, made of hydrogen and helium, and letting particle propagate and interact through the CRPropa Hadronic Interaction module. In Fig.~\ref{fig:HI_comp_1_1} is shown the case where only protons were injected, similar comparisons were already made in~\cite{HI_Dorner2025}, but only for $p-p$ interactions.

\section{CR distribution}

In Fig.~\ref{Fig:p_density}, we represented the proton CR distribution in the Galactic plane in different energy bins, obtained from our simulations, and normalized to $1$. As expected, increasing the energy yields smoother distributions, as CRs tend to follow trajectories closer to the ballistic case. Looking at the top panel, we can also note that at the lowest energies, CRs have some difficulties into reaching the most inner part of the MW before escaping or interacting. This results in a slight radial gradient in the CR density.

\section{Increasing statistics}

\begin{figure*}[!t]
    \centering

    % First figure: spans both columns
    \begin{minipage}{0.95\textwidth}
        \centering
        \includegraphics[width=0.7\linewidth]
        {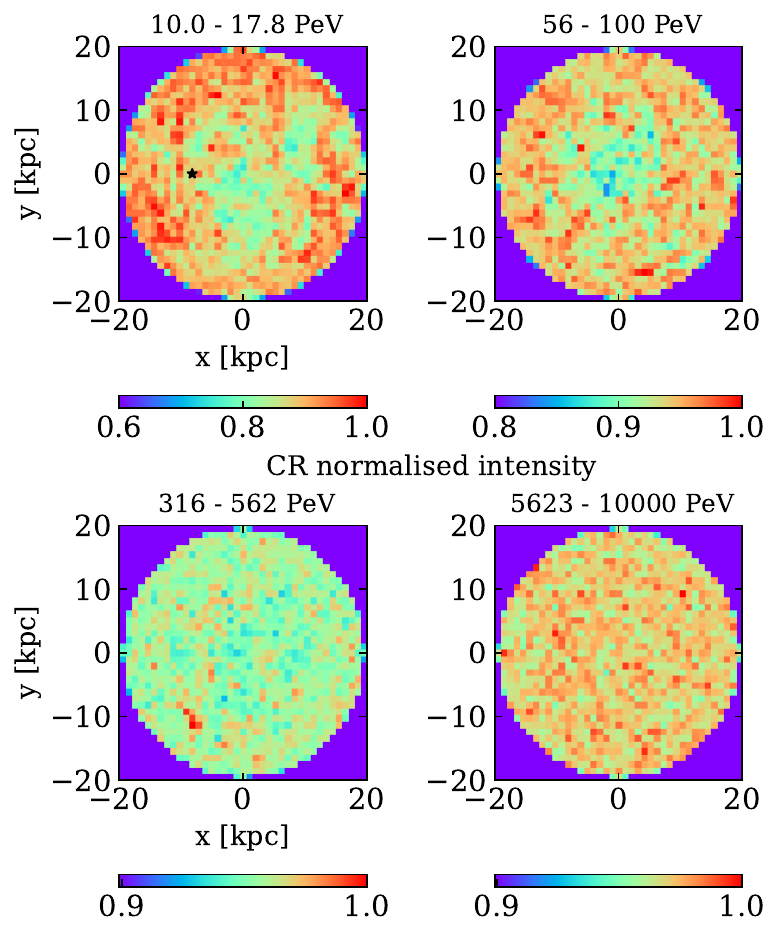}
        \caption{Intensity of CR protons in the Galactic plane for
        different energy bins. Values are normalised to 1 in each panel. The dark star in the upper left panel
        shows the position of Earth ($-8.178~\text{kpc}$ along the
        x-axis). Note that colorbars have different scales.}
        \label{Fig:p_density}
    \end{minipage}

    \vspace{0.5cm}

    % Second figure: separate caption and number
    \begin{minipage}{0.6\textwidth}
        \centering
        \includegraphics[width=\linewidth]{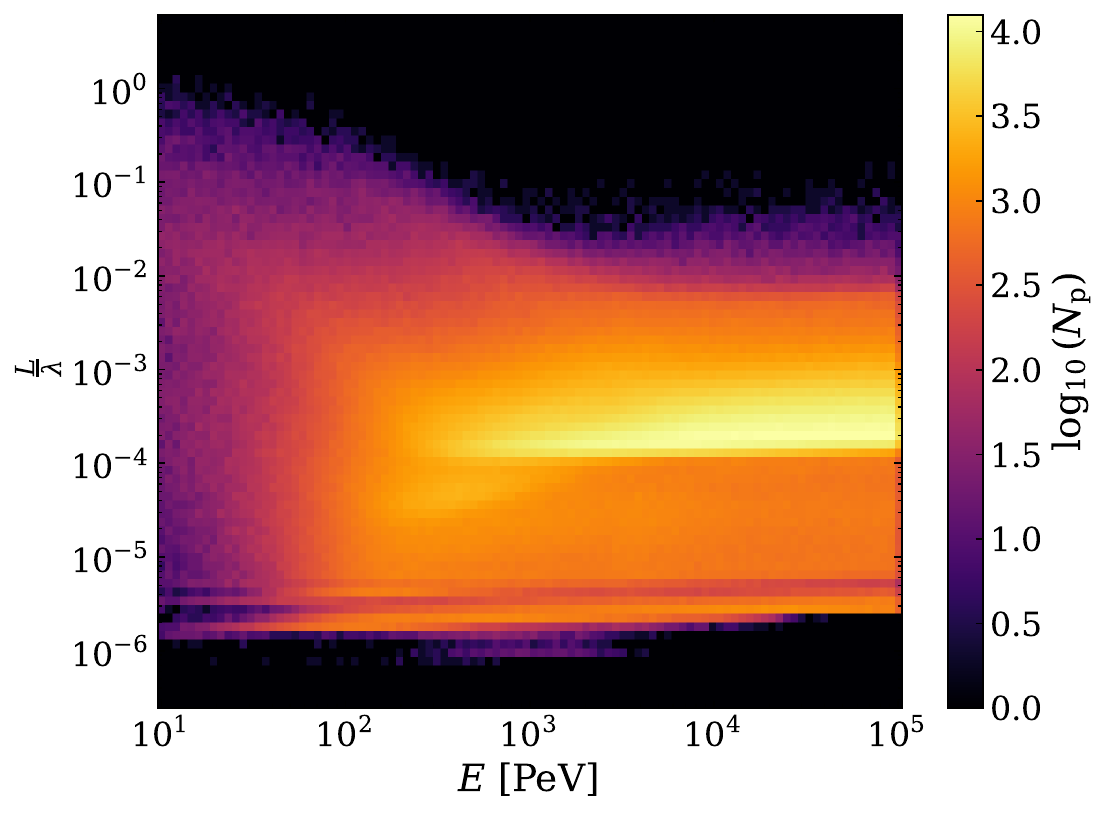}
        \caption{Ratio of the distance travelled by CRs $L$ multiplied by
        the gas density they encountered, and the mean free path $\lambda$ for
        a proton interacting with hydrogen gas in the MW.
        The energy displayed on the x-axis is the energy of the CR.}
        \label{fig:residence_time_protons}
    \end{minipage}

\end{figure*}

The mean free path for hadronic interactions of CRs with gas within the MW is much smaller than the distance most CRs travel within it, as showed in Fig \ref{fig:residence_time_protons}. This causes computational difficulties to sample a sufficient number of interactions. Because in our simulations particles produced close to the Earth have more chances to be detected at Earth, interactions in the vicinity of the Earth have to be properly sampled : we need to sample efficiently both the energy of the particles and the spatial distribution of the interactions. To increase the number of interactions, we used the CR density obtained through the multiple snapshots : combining it with the gas density, we simulated in each spatial and energy bins multiple hadronic interactions, before reweighting accordingly the produced secondaries. A similar procedure was performed for photopion production with the CMB and the EBL.

The intensity of neutrinos is computed assuming that their direction of emission from a given point is isotropic (which is the case if the incoming direction of CRs is uniformly distributed): every created neutrino is redirected so that it propagates towards Earth and hence contributes to the local intensity of neutrinos.

Our computation is similar for gamma rays, except that as described in the main text, they need to be properly propagated to account for absorption. In both cases the obtained intensities are then corrected by a geometric factor that accounts for this change of direction: the angular size of the observer as seen by the secondary when being emitted divided by $4 \pi$.

\end{document}